\documentclass{fairmeta}
\IfFileExists{libertine.sty}{}{%
  \ifPDFTeX
    \pdfmapfile{+newtx.map}
    \pdfmapfile{+qtm.map}
    \pdfmapfile{+qhv.map}
  \fi
  \usepackage{newtxtext,newtxmath}}
\usepackage{array,tabularx}
\usepackage{enumitem}

\usepackage{amsmath}
\usepackage{amssymb}
\usepackage{enumitem}
\usepackage{subcaption}
\usepackage{tikz}
\usetikzlibrary{positioning, arrows.meta, shadows, backgrounds, calc, shapes.geometric, fit}
\usepackage{balance}

\microtypesetup{expansion=false}
\newcommand{\ThreeKNEGain}{0.11}
\newcommand{\FiveKNEGain}{0.12}
\newcommand{\TenKNEGain}{0.19}
\newcommand{\ThreeKFlopReduction}{19.6\%}
\newcommand{\FiveKFlopReduction}{48.2\%}
\newcommand{\TenKFlopReduction}{69.6\%}
\newcommand{\LWRFrontierAlpha}{0.845}
\newcommand{\HSTUFrontierAlpha}{0.558}
\newcommand{\BaselineOneFrontierAlpha}{0.424}
\newcommand{\BaselineTwoFrontierAlpha}{0.599}
\newcommand{\LengthSharedNEFloor}{0.6789}

\newcommand{\SurfaceOneGlobal}{-0.08}
\newcommand{\SurfaceOneRegional}{-0.09}

\pdfcatalog{/Lang (en-US)}
\title{Read More, Reason Deeper: Latent Workspaces for Long-Sequence Recommendation}

\author[1]{Renzhi Wu}
\author[1]{Li Sheng}
\author[1]{Zikun Cui}
\author[1]{Junjie Yang}
\author[1]{Tai Guo}
\author[1]{Keke Zhai}
\author[1]{Yan Li}
\author[2]{Jerry Fu}
\author[1]{Nipun Mathur}
\author[1]{Sri Reddy}
\author[1]{Hong Li}
\affiliation[1]{Meta MRS}
\affiliation[2]{Facebook Discovery}

\hypersetup{
  pdftitle={Read More, Reason Deeper: Latent Workspaces for Long-Sequence Recommendation},
  pdfauthor={Anonymous Authors}
}

\abstract{
Long-sequence recommendation can improve along two axes: \emph{evidence
scaling} exposes longer histories, while \emph{reasoning scaling} adds layers
that interpret that evidence for a candidate. Existing models couple them:
sequence encoders propagate every added event through depth, whereas one-shot
selection prevents later layers from revisiting omitted evidence. We introduce
the Latent Workspace Ranker (LWR), which keeps typed evidence in persistent
banks and propagates a fixed-size candidate-local workspace. Each layer rereads
the banks from its evolving workspace and archives compact post-interaction
state, allowing history length and reasoning depth to grow independently.
We compare four architectures across various history--depth configurations using
evidence scaling at fixed depth, reasoning scaling at fixed 10K history, and
their joint compute-efficient frontier. LWR leads all families at long
histories, continues improving with reasoning compute, and converts additional
training compute into quality gains at the fastest fitted rate on the joint
frontier. At 10K history, it improves NE by \TenKNEGain{}\% over HSTU while
using \TenKFlopReduction{} fewer forward GFLOP/request. It also outperforms
existing baselines on two billion-scale ranking surfaces. These results show
that separating addressable evidence from propagated state makes reading more
and reasoning deeper independently controllable and jointly useful.
}

\date{\today}
\correspondence{Renzhi Wu at \email{renzhiwu@meta.com}}
\metadata[Keywords]{Long-sequence recommendation, Recommender systems}
\begin{document}

\maketitle

\section{Introduction}
Sequential recommendation models can typically be improved in two ways: More events
expose more evidence about a user's interests; more layers let a model
reinterpret that evidence as its representation of a candidate evolves. We
call these directions \emph{evidence scaling} and \emph{reasoning scaling}.
A useful architecture for long-sequence recommendation should control them independently and
benefit from both: reading more history should not enlarge the state that every
layer transforms, and adding layers should not cut off access to the original
history.

Yet current designs couple these axes. Sequence encoders keep event
representations active across layers
\citep{sasrec,bert4rec,hstu,cmsl}; lengthening history therefore expands work
throughout the stack. Target-aware pooling \citep{din,dien} and retrieval-based
rankers \citep{sim,ubr4ctr,sdim,twin,twinv2} bound active state by
summarizing evidence before feature interaction, but this makes evidence
selection effectively one-shot. Once an event is omitted, later layers cannot
recover it. This matters because the useful evidence can change during
reasoning: an early representation may match a candidate to a stable
preference, while a later one emphasizes the current session. Existing designs
thus face a structural choice: carry more evidence through every layer, or let
deeper reasoning operate on evidence chosen only once.

\noindent\textbf{Latent workspaces decouple the two axes.}
We introduce the \emph{Latent Workspace Ranker} (LWR), which separates what the
model can read from what it must carry. Typed history, request/user context,
and matched events remain in persistent external banks. Candidate features
initialize a fixed-size workspace that alone moves through the stack. Evidence
scaling increases bank cardinality and read length, whereas reasoning scaling
increases the number of workspace layers. These are separate structural
controls: changing either does not enlarge the representation governed by the
other.

\begin{figure*}[t]
\centering
\includegraphics[width=\textwidth]{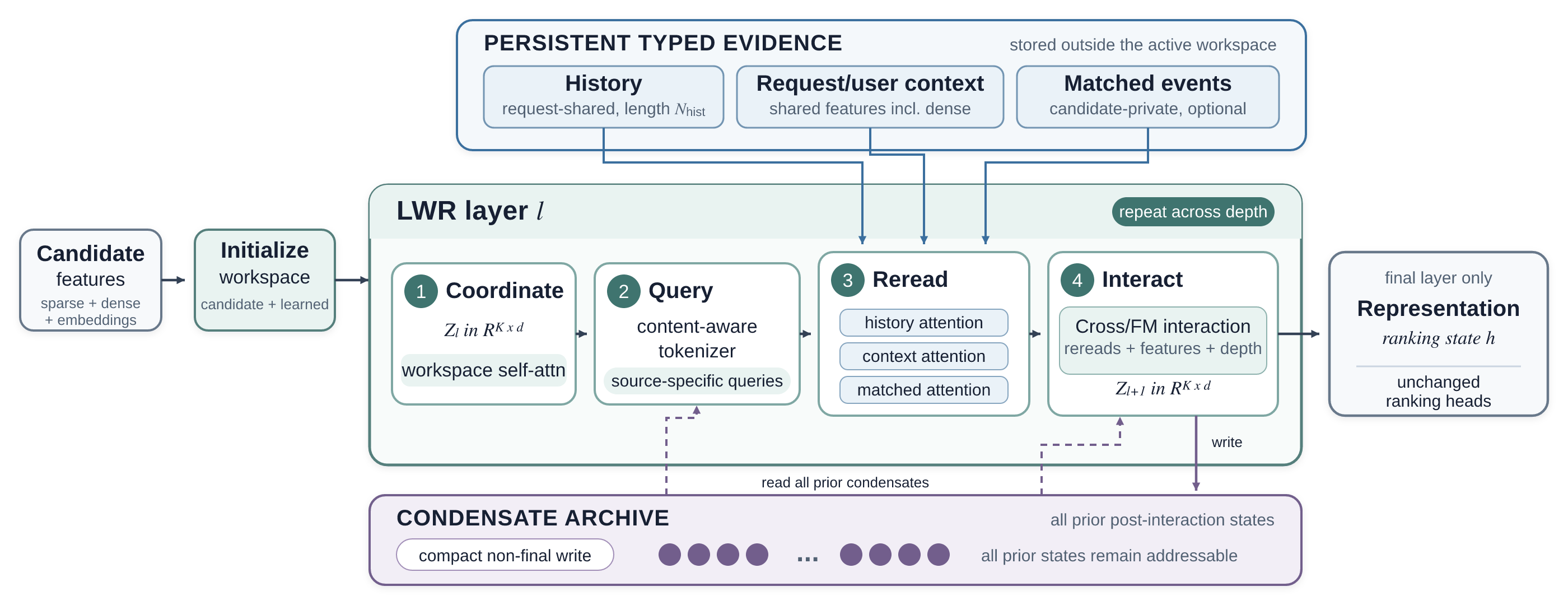}
\caption{LWR decouples evidence scaling from reasoning scaling. Persistent typed
banks retain history and request/user features, including dense features;
candidate features seed a fixed-size workspace that moves through repeated
layers. Each layer coordinates, queries, rereads, and interacts; non-final
layers write compact state for subsequent reasoning.}
\label{fig:architecture}
\end{figure*}

This separation is useful only if later layers can use the retained evidence
differently. Each LWR layer therefore \emph{Coordinates} its workspace, uses
the result to \emph{Query} each bank, \emph{Rereads} the returned evidence, and
\emph{Interacts} the responses with ranking features
(Figure~\ref{fig:architecture}). Non-final layers write a few
\emph{condensate} tokens to an archive. Because both the workspace and archive
evolve, a later layer can formulate a candidate-specific follow-up read from
earlier results while still accessing the original evidence. Longer histories
provide more addressable evidence; deeper stacks ask better-informed questions
of it.

\noindent\textbf{Three curves expose two-axis scaling.}
The central test is whether the separated axes produce useful gains rather
than merely configurable knobs. We evaluate four architectures on a grid of
history lengths and reasoning depths, then read that grid in three ways.
The evidence curve holds depth fixed and asks whether longer history helps.
The reasoning curve holds history fixed and asks whether added reasoning
continues to pay off relative to its training compute. The joint curve removes
both restrictions: among all history--depth combinations, it traces the
configurations that are not beaten in both compute and NE. It therefore asks
how well each architecture allocates a compute budget between reading more and
reasoning deeper, rather than forcing both axes to increase in lockstep.

LWR leads all four families in the long-history regime, continues improving
throughout the fixed-history reasoning sweep, and has the steepest joint
compute-frontier trend. Its reducible normalized entropy (NE), the
normalized-log-loss error remaining above a fitted floor, falls fastest as
training compute grows along that frontier.
At 10K history, LWR achieves a statistically significant \TenKNEGain{}\% NE
improvement over HSTU while using
\TenKFlopReduction{} fewer forward GFLOP/request.
In addition, LWR improves critical-task NE by 0.08--0.09\% over the existing baseline on Surface 1 and improves NE by 0.15--0.30\% over HSTU \citep{hstu} on Surface 2.

Our contributions are:
\begin{itemize}[leftmargin=*]
  \item \textbf{An architecture for decoupled scaling:} LWR keeps evidence in
  persistent typed banks while a fixed-size candidate-local workspace moves
  through depth. Repeated reads preserve access to original evidence, and a
  condensate archive carries compact intermediate reasoning across layers.
  \item \textbf{Three complementary scaling curves:} We isolate evidence
  scaling at fixed depth and reasoning scaling at fixed history, then define
  joint scaling as the compute-efficient envelope of the full history--depth
  grid. LWR benefits strongly from longer histories, continues improving with
  deeper reasoning, and has the steepest joint-frontier trend.
  \item \textbf{Billion-scale quality and compute efficiency:} LWR outperforms existing baselines on two ranking
  surfaces.
  At 10K history it improves NE by \TenKNEGain{}\% over HSTU with
  \TenKFlopReduction{} fewer forward GFLOP/request.
\end{itemize}

\section{Related Work}

Sequential recommenders encode history with recurrent \citep{dien},
self-attentive \citep{sasrec,bert4rec}, or generative sequence models
\citep{hstu}. Long-history methods restrict the active evidence through
external memory \citep{mimn}, target-aware activation \citep{din}, retrieval
\citep{sim,ubr4ctr,twin,twinv2,rankgraph,rankgraph_2}, or sampling \citep{sdim};
efficient long-context models instead reduce sequence-processing cost through
sparsity or IO-aware kernels
\citep{transformerxl,longformer,bigbird,reformer,flashattention}. These methods
make longer histories practical, but generally do not separate the evidence
retained from the state propagated through later ranking interactions.

Bounded-latent models provide complementary ingredients: Set Transformer uses
inducing points \citep{settransformer}, Perceiver alternates cross-attention and
latent processing \citep{perceiver,perceiverio}, and memory networks repeatedly
read external state \citep{e2ememory}. LWR adapts these ideas to ranking with
source-specific queries, explicit FM/deep-cross interactions
\citep{widedeep,fm,deepfm,xdeepfm,dcn,autoint,fibinet}, and compressed
post-interaction state shared across layers. The contribution is their
composition for decoupled evidence and reasoning scaling, together with its
long-sequence evaluation.

Most directly, SlimPer \citep{slimper} queries raw user-side tokens while
iteratively refining a compact knowledge base. LWR instead separates
persistent typed evidence, a candidate-local workspace, and an all-prior
archive of compressed post-interaction states.

\section{Latent Workspace Ranker}

\subsection{Architecture overview}

LWR separates persistent evidence from compact propagated state. For request
$r$ and candidate $i$, let $x_{r,i}$ be candidate context, $f_{r,i}$ a dense
interaction projection, and
$\{S_r^{\rm hist},S_r^{\rm ctx},S_{r,i}^{\rm sim}\}$ the typed evidence
banks. We use $\mathcal S=\{\mathrm{hist},\mathrm{ctx},\mathrm{sim}\}$ to
index them and $S^s$ when request and candidate indices are clear. History and
request context are request-shared; matched events are candidate-private.

At layer $l$, the propagated state is a candidate-local workspace
$Z_l\in\mathbf R^{K\times d}$ and an archive
$C_{<l}=[c_0;\ldots;c_{l-1}]$ of completed layer states, with
$C_{<0}=\emptyset$. Here $K$ is the workspace slot count, $d$ the hidden
width, and each of the $D$ layers maps $Z_l$ to $Z_{l+1}$.
Figure~\ref{fig:architecture} shows the four steps: \emph{Coordinate} the
workspace, \emph{Query} each source, \emph{Reread} its evidence, and
\emph{Interact} the responses with ranking features. Each source receives
$r_s$ queries, and every non-final layer appends $q_c$ condensate tokens to the
archive. The following subsections define these operations.

The same structure exposes two scaling controls. Evidence scaling changes
$N_s=|S^s|$, while reasoning scaling changes $D$. Keeping $K$, $r_s$, and
$q_c$ bounded prevents either axis from enlarging the state controlled by the
other. Source reads still scale with $N_s$ and repeat across $D$; LWR decouples
state growth, not total computation.

\subsection{Coordinate: evolve a compact workspace}

The initial workspace combines candidate-derived and learned slots:
\begin{align}
 Z_0&=[Q_0;M_0]=[\phi_{\rm cand}(x_{r,i});\Theta_M],
 &K&=n_Q+n_M, \nonumber\\
 \bar Z_l&=\operatorname{WSA}_l(\operatorname{Norm}_l(Z_l)).
\label{eq:coordinate}
\end{align}
Here $\phi_{\rm cand}$ maps candidate context to $n_Q$ slots,
$\Theta_M\in\mathbf R^{n_M\times d}$ contains $n_M$ learned slots, and
$Q_0=\phi_{\rm cand}(x_{r,i})$, $M_0=\Theta_M$. The learned bank is broadcast
to every candidate and supplies content-independent working capacity.
$\operatorname{Norm}_l$ and $\operatorname{WSA}_l$ denote the layer's
normalization and workspace self-attention. $Q$ and $M$ describe
initialization provenance only; subsequent operations mix all slots into
mutable state. Unlike a fixed multi-interest bank \citep{mind}, learned slots
need not retain static semantic roles.
Coordinate applies normalized, dense, non-causal self-attention over the $K$
slots of one candidate.

\subsection{Query: form source-specific requests}

The tokenizer $T_l$ converts the coordinated workspace $\bar Z_l$, prior
condensates $C_{<l}$, and layer-specific learned prior tokens $P_l$ into one
bounded query bank $U_l^s\in\mathbf R^{r_s\times d}$ per source $s$:
\begin{equation}
 U_l=T_l([\bar Z_l;C_{<l};P_l]),\qquad
 \{U_l^s\}_s=\operatorname{Split}(U_l).
\label{eq:query}
\end{equation}
Here $U_l$ is the combined query bank, and $\operatorname{Split}$ partitions
it into the prescribed $r_s$ queries for each source.
Because the workspace and archive change with depth,
so do the queries: an early layer can ask a broad history question, while a
later layer can ask a candidate-specific follow-up informed by completed
interactions. Query is the bridge between the two scaling axes: larger banks
provide more addressable evidence, while added layers generate new requests
from the evolving reasoning state.

Separate banks preserve source identity instead of forcing heterogeneous
evidence into one undifferentiated query. Query construction retains a static
token-mixing arm and adds a content-dependent arm whose output projection is
initialized at zero. The initial function is therefore preserved while the
adaptive path receives gradients from the first update.

\subsection{Reread: access persistent typed evidence}

Each $U_l^s$ rereads source $S^s$ through cross-attention. Packed history and
request/user context are request-scoped, while matched events are
candidate-scoped. Grouped offsets let all candidates address their request's
shared banks without copying them; candidate offsets keep private reads from
crossing examples. The original typed sources remain available at every layer,
so later computation need not accept an irrevocable early summary:
$
 R_l^s=\operatorname{Attn}(U_l^s,S^s,S^s)
$
where $\operatorname{Attn}$ denotes multi-head cross-attention and
$R_l^s\in\mathbf R^{r_s\times d}$ is the response from source $s$.

This step is related to multi-hop external memory \citep{e2ememory} and iterative
latent reads \citep{perceiver,perceiverio}, but it retains source ownership and
places an explicit ranking interaction after retrieval. Bounded queries make
the propagated state independent of source length, not the read itself: a
source of length $N_s$ with $r_s$ queries still contributes
$O(r_sN_sd)$ attention work.

\subsection{Interact: update ranking state}

The block $F_l$ is a ranking-specific Cross/FM interaction over the coordinated
workspace, retrieved source responses, earlier condensates, and dense feature:
$
 Z_{l+1}=F_l([\bar Z_l;\{R_l^s\}_s;C_{<l}],f_{r,i}).
$
This keeps evidence access distinct from explicit feature crossing, following
the FM and deep-cross lineage \citep{fm,deepfm,xdeepfm,dcn,autoint,fibinet}. Its
output is again $Z_{l+1}\in\mathbf R^{K\times d}$, so deep candidate reasoning
does not inherit source cardinality.
The final layer emits a fixed-width representation to the PLE multi-task module and its task heads.

\begin{figure*}[t]
\centering
\includegraphics[width=\textwidth]{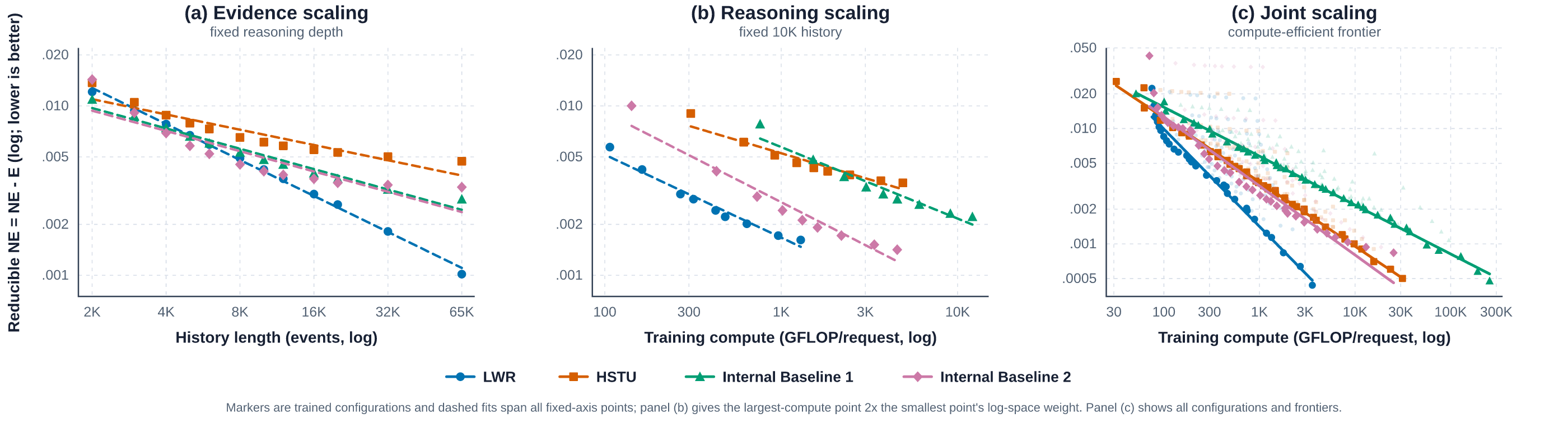}
\caption{Evidence, reasoning, and joint scaling; lower is better.
Figure~\ref{fig:scaling}(a) varies history from 2K to 65K at fixed depth, and
Figure~\ref{fig:scaling}(b) varies depth at fixed 10K history against total
training compute. Both plot reducible NE, $\mathrm{NE}-E$, with a shared floor
$E=\LengthSharedNEFloor{}$, preserving raw-NE ordering.
Figure~\ref{fig:scaling}(c) retains the compute-efficient frontiers from all
468 history--depth configurations and fits an architecture-specific floor.
Markers denote trained configurations and curves denote power-law fits; all
axes are logarithmic.}
\label{fig:scaling}
\end{figure*}

\subsection{Stacking layers through a condensate archive}

Every non-final layer applies a learned compressor
$\kappa_l:\mathbf R^{K\times d}\rightarrow\mathbf R^{q_c\times d}$ to its
\emph{post-interaction} workspace and appends the resulting $q_c$ tokens to
the archive:
\begin{equation}
 c_l=\kappa_l(Z_{l+1}),\qquad
 C_{<l+1}=[C_{<l};c_l]\quad(l<D-1).
\label{eq:condensate}
\end{equation}
A condensate therefore records how
that layer combined candidate state, retrieved evidence, and ranking features;
it is not another copy of source tokens. Thus
$c_l\in\mathbf R^{q_c\times d}$ and
$C_{<l}\in\mathbf R^{lq_c\times d}$. Layer $l$ receives all earlier
condensates $C_{<l}$ in both Query and Interact. It can reread an original
source, consult a compact record of prior decisions, or combine both paths.

This all-prior connection gives later layers direct access to every completed
stage without propagating full intermediate activations. It differs from
uncompressed dense concatenation \citep{densenet} and temporal compression
across sequence segments \citep{compressivetransformer}: LWR compresses state
across network depth. Reasoning scaling increases $D$; if each layer writes
$q_c$ tokens, layer $l$ reads $lq_c$ archive tokens, and the stored
candidate-local state is $O((K+Dq_c)d)$. Evidence scaling changes the source
lengths used only in Reread; the large banks remain external and can stay
request-shared.

The four-step layer and archive together create two recurrent paths: original
evidence is revisited across sequence length, while compressed decisions are
revisited across network depth.

\section{Experiments}

\subsection{Setup and metrics}

We first compare LWR with existing reference rankers on two surfaces, then
use one history--depth grid to isolate evidence scaling, reasoning scaling, and
their best joint use of compute.

\noindent\textbf{Data and hardware.}
Each model trains on three consecutive days of online data (30--50 billion
examples, depending on the surface) and evaluates on the following day. Runs
use mixed precision on 128 NVIDIA H100 GPUs on an internal training platform
with customized memory and throughput kernels.

\noindent\textbf{Settings.}
Overall comparisons use 3K histories, with HSTU as the Surface 2 reference and
a existing baseline as the Surface 1 reference.
LWR uses five layers, 76 workspace slots
, 24 queries per typed source, and four
condensates per non-final layer; its 2048-dimensional output feeds the unchanged
PLE module.

\noindent\textbf{Scaling protocol.}
Evidence scaling uses 12 history lengths from 2K to 65K at base depth.
Reasoning scaling uses nine depths at fixed 10K history and plots each run at
its measured total training compute. Joint scaling takes the nondominated
lower envelope of all 117 configurations per family: a point is retained only
if no other point has both no worse NE and no greater compute, with one strict
improvement. It therefore measures the best allocation of compute between
history and depth.

\noindent\textbf{Metrics.}
Normalized entropy (NE)~\citep{he2014practical} is binary log loss divided by empirical label entropy;
lower is better. We report NE reductions as
$100(\mathrm{NE}_{\rm base}-\mathrm{NE}_{\rm LWR})$, and improvements above
0.03\% are statistically significant under the surface-level protocol.
Surface 1 instead reports critical-task p-diff, where negative is better; the
two surface metrics are not combined. Forward GFLOP/request measures the 3K,
5K, and 10K headline slice, while total forward-plus-backward training compute
defines the reasoning and joint curves. Figure~\ref{fig:scaling} reports
reducible NE, $\mathrm{NE}-E$: Figures~\ref{fig:scaling}(a) and (b) share one
floor to preserve raw-NE ordering, while Figure~\ref{fig:scaling}(c) fits one
floor per architecture to compare frontier slopes.

\subsection{Overall results on two ranking surfaces}
\label{sec:overall-results}

\begin{table*}[t]
\centering
\small
\renewcommand{\arraystretch}{1.04}
\setlength{\tabcolsep}{5.5pt}
\begin{tabular}{lrrr}
\toprule
Surface and metric & Reference & LWR & Difference (\%) \\
\midrule
Surface 2 task A train NE & 0.6815 & 0.6796 & $-0.19\%$ \\
Surface 2 task A eval NE  & 0.6841 & 0.6826 & $-0.15\%$ \\
Surface 2 task B train NE & 0.6297 & 0.6267 & $-0.30\%$ \\
Surface 2 task B eval NE  & 0.6276 & 0.6251 & $-0.25\%$ \\
Surface 1 eval global NE diff & 0 & \SurfaceOneGlobal{}\% & \SurfaceOneGlobal{}\% \\
Surface 1 eval regional NE diff & 0 & \SurfaceOneRegional{}\% & \SurfaceOneRegional{}\% \\
\bottomrule
\end{tabular}
\caption{Overall quality at 3K history. Surface 2 compares LWR with HSTU
\citep{hstu}; Surface 1 reports critical-task p-diff relative to a existing
baseline. Lower is better, and metrics are not comparable across surfaces.}
\label{tab:overall}
\end{table*}

Table~\ref{tab:overall} shows consistent gains across both surfaces. LWR
improves Surface 2 train and evaluation NE by 0.15--0.30\%, all above the
0.03\% significance threshold, and yields NE diff of \SurfaceOneGlobal{}\%
globally and \SurfaceOneRegional{}\% regionally on Surface 1. The metrics are
surface-specific, so we do not pool them.


\subsection{Evidence, reasoning, and joint scaling}

\noindent\textbf{Evidence scaling: read more at fixed depth.}
Figure~\ref{fig:scaling}(a) isolates evidence scaling by fixing the base
reasoning depth. At 3K, 5K, and 10K history, LWR improves NE over HSTU by
\ThreeKNEGain{}\%,
\FiveKNEGain{}\%, and \TenKNEGain{}\% while reducing measured forward
GFLOP/request by \ThreeKFlopReduction{}, \FiveKFlopReduction{}, and
\TenKFlopReduction{}; all three NE gains exceed the 0.03\% significance
threshold. At 10K, Internal Baseline 2 is only 0.01\% better, below the
threshold, but uses 136.28 rather than 52.69 GFLOP/request. Across the full
2K--65K sweep, LWR starts behind Internal Baseline 1 and leads all four
families from 12K onward, showing that its advantage grows with available
evidence.

\noindent\textbf{Reasoning scaling: reason deeper at fixed evidence.}
Figure~\ref{fig:scaling}(b) isolates reasoning scaling by holding history at
10,240 events and varying relative depth from $0.5\times$ to $8\times$. All
four families improve monotonically, and LWR improves NE by 0.41\% as deeper
workspace layers extract more value from unchanged evidence. Total training
compute, rather than nominal depth, makes this a quality-per-reasoning-compute
comparison. At $8\times$ depth, LWR outperforms HSTU and Internal Baseline 1;
Internal Baseline 2 is 0.02\% better, below the significance threshold, but
uses 3.5 times as much training compute.

\noindent\textbf{Joint scaling: spend compute on either axis.}
Figure~\ref{fig:scaling}(c) takes the compute-efficient envelope of every
history--depth combination, measuring the best quality when compute may be
spent on either axis. Following neural scaling studies
\citep{hestness,kaplan,hoffmann}, we fit
$\mathrm{NE}=E_m+A_mC^{-\alpha_m}$ to each frontier, where $\alpha_m$ measures
how quickly reducible NE falls under the best allocation. LWR has the steepest
frontier exponent:
$\alpha=\LWRFrontierAlpha{}$, versus \HSTUFrontierAlpha{},
\BaselineOneFrontierAlpha{}, and \BaselineTwoFrontierAlpha{} for HSTU and
Internal Baselines 1 and 2. Figure~\ref{fig:scaling}(c) fits $E_m$ separately
for each family, so vertical position represents reducible scaling behavior,
not raw-NE ordering. Together, the three curves show that LWR benefits from
more evidence, deeper reasoning, and their joint allocation.

\section{Conclusion}

Latent Workspace Ranker decouples evidence scaling from reasoning scaling:
persistent typed banks retain long histories, while a compact workspace
repeatedly reads them and carries state across layers. It outperforms HSTU and
a existing baseline on two billion-scale ranking surfaces. At fixed depth,
LWR leads all evaluated families from 12K through 65K history; at fixed
history, it continues improving with reasoning compute; and across all 468
history--depth configurations, it has the steepest fitted compute-efficient
frontier.
Together, these results show that long-sequence recommenders can read more and
reason deeper without coupling both forms of capacity to the same expanding
sequence state.

\label{page:references}
\bibliographystyle{ACM-Reference-Format}
\bibliography{references}

\end{document}